\documentclass[12pt]{iopart}
\usepackage[pdftex]{graphicx}
\usepackage{iopams}
\usepackage{lineno}
\usepackage{bm}
\usepackage{siunitx}
\usepackage{soul}
\usepackage{url}

\newcommand{\code}[1]{{\color{mydarkgrey} \texttt{\lstinline|#1|}}}

\begin{document}

\title[GAN for MC simulations]{Generative Adversarial Networks (GAN) for compact beam source modelling in Monte Carlo simulations}



\author{D.~Sarrut\textsuperscript{1}, N.~Krah\textsuperscript{1,2}, JM.~L\'etang\textsuperscript{1}}

\address{$^1$ Universit\'e de Lyon, CREATIS; CNRS UMR5220; Inserm U1044;
  INSA-Lyon; Universit\'e Lyon 1; Centre L\'eon B\'erard, France.\\
  $^2$ University of Lyon, CNRS/IN2P3, Institute of Nuclear Physics, UMR 5822, Villeurbanne, France}

\ead{david.sarrut@creatis.insa-lyon.fr}

\newcommand\MC{Monte Carlo}
\newcommand\Tc{$^{99m}$Tc}
\newcommand\Indium{$^{111}$In}
\newcommand\Iodine{$^{131}$I}
\newcommand\Lutetium{$^{177}$Lu}
\newcommand\X{\textbf{X}}
\newcommand\Z{\textbf{Z}}
\bmdefine{\bX}{X}
\bmdefine{\bZ}{Z}
\bmdefine{\bY}{Y}
\bmdefine{\bx}{x}
\bmdefine{\bz}{z}
\bmdefine{\bmu}{\mu}
\bmdefine{\bSigma}{\Sigma}
\bmdefine{\bsigma}{\sigma}
\bmdefine{\btheta}{\theta}
\bmdefine{\bphi}{\phi}


\begin{abstract}
  
  \noindent A method is proposed and evaluated to model large and
inconvenient phase space files used in Monte Carlo simulations
by a compact Generative Adversarial Network (GAN). The GAN is trained based on
a phase space dataset to create a neural network, called Generator (G),
allowing G to mimic the multidimensional data distribution of the
phase space. At the end of the training process, G is stored with about 0.5
million weights, around 10\,MB, instead of few GB of the initial
file. Particles are then generated with G to replace the phase space dataset.

This concept is applied to beam models from linear accelerators (linacs) and from
brachytherapy seed models. Simulations using particles from the
reference phase space on one hand and those generated by the GAN on the other
hand were compared. 3D distributions of deposited energy 
obtained from source distributions generated by the GAN were close to the reference ones, with less than 1\%
of voxel-by-voxel relative difference. Sharp parts such as the
brachytherapy emission lines in the energy spectra were not perfectly modeled by the
GAN. Detailed statistical properties and limitations of the
GAN-generated particles still require further investigation, but the
proposed exploratory approach is already promising and paves the way
for a wide range of applications.


\end{abstract}




\section{Introduction}

Monte Carlo simulations are widely used to characterize sources of
particles such as those of linac photon/electron beams, X-ray tubes,
proton beam nozzles, brachytherapy radionuclide seeds, particles
emerging from a voxelised patient geometry (to simulate a nuclear
imaging process), etc. The computation time to perform such
simulations is generally high and phase space files have been
acknowledged as a necessary means to avoid repeated and redundant execution of part of the simulation. A typical example is the dose calculation in a patient CT image where the simulation is split into two parts~\cite{Andreo2018}. A first, detailed, Monte Carlo simulation is performed to transport particles through the accelerator treatment head elements (primary collimation, flattening filter, monitor chambers, mirrors, secondary collimation, etc.), up to a virtual plane. The properties (energy, position, direction) of all particles reaching the plane are stored in the phase space file and depend on the detailed properties of the treatment head components, such as its shape and materials. A second simulation tracks particles from the phase space plane through the multi-leaf collimator and the patient CT image to estimate the absorbed dose distribution. 

However, phase space files are typically up to several tens of
gigabytes large and inconvenient to use efficiently.  Statistical limitations may also arise when more particles are required than stored in the phase space file. Several virtual source
models for linac beam modelling have been proposed in the
literature. For example in~\cite{Fix2001, Grevillot2011}, the authors
describe the statistical properties of the phase space distribution by analytical
functions, by evaluating the dependence of the parameters and by
adapted sampling procedures.  Chabert et al.~\cite{Chabert2016} used
4D correlated histograms with different adaptive binning schemes to
represent an Elekta Synergy 6 MV photon beam.  Recently, Brualla et
al.~\cite{Brualla2019} proposed a method to extract light-weight
spectra from large phase space files. This method may be sufficient for some
applications, but may neglect correlations between energy, position
and direction and require adaptive binning. Overall, the proposed
methods were useful, but are specific to a simulation type and may not
be applied in applications other than linac simulations. In this work,
we propose the use of GAN as a generic beam modelling technique.





Generative Adversarial Networks (GAN) were recently
reported~\cite{Goodfellow2014, Creswell2018} as deep neural network
architectures which allow mimicking a distribution of multidimensional
data and have gained large popularity due to their success in
realistic image synthesis. GAN learn representations of a training
dataset by implicitly modeling high-dimensional distributions of
data. Once trained, the resulting model is a neural network called
Generator $G$ that produces an element $\bx$ that is supposed to
belong to the underlying probability distribution of the training
data. Based on this principle, several variants have been proposed such
as the Wasserstein GAN (WGAN)~\cite{Arjovsky2017} that will be used in
this work. GAN and derivatives are still emerging concepts and an
active field of research.



The idea to use machine learning techniques for Monte Carlo
simulations and high energy physics is not new. Several works
have been reported such as the use of neural networks in
condensed matter physics~\cite{Carrasquilla2017,Shen2018} or the use of
GAN~\cite{Paganini2018} for fast simulation of particle showers in
electro-magnetic calorimeters. However, to our knowledge, GANs have never
been proposed for representing large phase space distributions.










\section{Materials \& Methods}

\subsection{Training data}

A phase space dataset generated by a MC simulation contains a set of particles
described by position, direction cosines, energy, statistical weight,
etc. The dataset is intended to be as large as possible in order to be
representative of the underlying probability distribution of the
particle source, denoted as $p_{\mathrm{real}}$. A phase space file is
an extensive finite description of multidimensional data and thus may
suffer from residual uncertainty known as latent
variance~\cite{Sempau2001}. It is used as a training dataset with
samples $\bx \in \mathbb{R}^d$ of dimension $d$. In practice, $d$ is
typically equal to 7, with 3 parameters for the particle position, 3
for the direction and 1 for the energy.






\subsection{GAN optimisation}

In the following, we
first introduce the general concept behind a GAN and then summarize the
aspects specific to a WGAN, a special category of GAN used in this work. 

The goal is to train a generative function $G$ that models a
distribution $p_{\btheta}$. Parameters $\btheta$ are related to the
distribution model approximating the target distribution
$p_{\mathrm{real}}$ only known by samples from a training
dataset~\cite{Goodfellow2014}. The neural network architecture is
composed of two multilayer perceptrons, $D$ and $G$, competing 
against each other, hence the term “adversarial”. The \emph{generator}
$G(\bz ; \btheta_{G})$ is trained to produce samples distributed
similarly as the data distribution of $\bx$. It takes $\bz$ as input,
sampled from a simple multidimensional normal prior distribution,
$\mathcal{N}^d(0,1)$, and produces a sample $\bx$ as if it were drawn
from $p_{\textrm{real}}$.  The parameters $\btheta_{G}$ are the
weights of the network $G$. The \emph{discriminator} $D(\bx ;
\btheta_{D})$ is trained to distinguish between samples from the real data
distribution and those generated by $G$. It takes $\bx$ as input and
yield a single scalar as output that represents the probability of $\bx$
coming from the real data rather than from the generator. $D$ is
trained to maximize the probability of correctly identifying samples
from the training data as real and those generated by $G$ as fake. The
parameters $\btheta_{D}$ are the weights of the network $D$.

The GAN training process is a zero-sum non-cooperative game which
converges when the discriminator and the generator reach Nash
equilibrium~\cite{Fedus2018}. At such an equilibrium, one
player (neural network) will not change its action (weights)
regardless of what the opponent (the other network) may do. In the
conventional GAN formulation~\cite{Goodfellow2014}, the considered
cost function was the Binary Cross Entropy (BCE) for both $G$ and
$D$. $\textrm{BCE}(p,q)$ between two distributions $p$ and $q$ is
related to the Kullback-Leibler divergence which measures the
performance of a classification model whose output is a probability
value between 0 and 1. 
The loss function of a GAN
quantifies the similarity between the data distribution generated by
$G$ and the real sample distribution and it has been shown that this 
corresponds to the Jensen-Shannon divergence
(JSD) when the discriminator is optimal~\cite{Goodfellow2014}. JSD is
a symmetric and smooth version of the Kullback-Leibler divergence.

In practice, we found a conventional GAN difficult to train and
subject to mode collapse~\cite{Arjovsky2017}. Instead, we 
used WGAN, proposed by Arjovsky et
al.~\cite{Arjovsky2017}, which uses the Wasserstein (or Earth Mover's) distance as an
alternative loss function. The Wasserstein
distance between two distributions $p$ and $q$ is the cost of the
optimal transport needed to deform $p$ into $q$. It has been shown that it
helps stabilizing the learning process, because it is less subject to
vanishing gradients than a conventional GAN. In practice, there are few
changes compared to the original GAN. First, the loss functions become the
following expressions:

\begin{eqnarray}
  J_{D}\left( \btheta_{D}, \btheta_{G} \right) & = & \mathbb{E}_{\bz} \left[D(G(\bz))\right] -
  \mathbb{E}_{\bx} \left[D(\bx)\right]  \label{eq:JDW} \\ J_{G}\left( \btheta_{D}, \btheta_{G} \right) &
  = & -\mathbb{E}_{\bz} [D(G(\bz))] \label{eq:JGW}
\end{eqnarray}

Second, after every gradient update, the weights $\btheta_{D}$ are
clamped to a small fixed range (e.g.~$[-0.01, 0.01]$) in order to
constrain weights to a compact space. Finally, the
authors~\cite{Arjovsky2017} also recommend using the RMSProp
optimizer~\cite{Tieleman2012} instead of the conventional Adam
optimizer~\cite{Kingma2014} because the latter uses momentum processes that may
cause instability in the model training. In a WGAN, $D$ does not act
as a explicit discriminator, but is a helper for approximating the
Wasserstein metric between real and generated data distributions. $D$
is called the ``critic''. The training is no more performed until Nash
equilibrium, but until loss convergence.

\subsection{GAN architecture and parameters}
\label{sec:parameters}

A GAN requires setting several interconnected hyperparameters that
influence results in different ways. We describe here the optimal set
of parameters we empirically found. We comment on their influence we
observed in the result section. The architecture of both $D$ and $G$
was the following: we used $H=400$ neurons in each of the 3 hidden
linear fully connected layers. The values were set empirically, based
on experimental results. As advocated in~\cite{Arjovsky2017}, the
activation function was a Rectified Linear Unit (ReLu) $r(x) =
\max(0,x)$ for all layers except for the last one of $G$ where instead
a sigmoid function was used. We set the dimension of $\bz$ to 6. The
total number of weights was around $\num{5e5}$ for both $D$ and
$G$. The learning rate was chosen empirically and set to
$\num{e-5}$. Stochastic batches of $\num{e4}$ samples were used at
each iteration. The discriminator was updated more frequently than the
generator, four times per iteration versus once, as advocated
in~\cite{Arjovsky2017}. We set the number of iterations (or
\emph{epochs} in the deep learning community) to $80\,000$.






\subsection{Implementation}

All simulations were implemented in Gate version
8.0~\cite{Sarrut2014}, using Geant4 version
10.3~\cite{Allison2016}. Neural network operations (training, samples
generation) were implemented in Python with the PyTorch
framework~\cite{Paszke2017} using CUDA GPU acceleration. Once the
network is trained, it can be read and used within Gate/Geant4 thanks
to a newly developed Gate module exploiting the PyTorch C++
API~\cite{pytorch-cpp2019}. Source code is
open-source and will be available in the next Gate
release. Computations were performed on an Intel Xeon CPU E5-2640 v4 @
2.40GHz and an NVIDIA Titan Xp (GP102-450-A1) with 12 GB memory.

\subsection{Experiments}

\newcommand{\edep}{D}

The proposed method was evaluated with phase space files of two linear
accelerators provided by IAEA~\cite{Capote2006} and of brachytherapy
$^{125}$I seeds~\cite{Thiam2008,Sarrut2014}. The linacs are a 6 MV
photon Elekta and a Cyberknife with IRIS collimator of 60 mm. For the
linac phase space files, the particles were parameterized by energy
$E$, position $x,y,z$, and direction cosines $dx, dy, dz$. Both phase
space files were recorded in a plane, so the $z$ coordinate was
constant and ignored in the following, leading to 6 dimensions.  The
phase space distribution are available as two files (see
table~\ref{tab:datasets}): one file was used to train the GAN
(PHSP$_1$); the other file was used for evaluation (PHSP$_2$). To
generate a purely continuous energy distribution, the phase space
datasets were pre-processed to remove the 511\,keV peak. The phase
space files of the $^{125}$I brachytherapy seed (59.49\,days half
life, maximum 35\,keV) were obtained from simulations of a seed
capsule model composed of a double-wall made of titanium surrounding a
tungsten x-ray marker coated with an organic carbon
layer~\cite{Sarrut2014} (The Best Medical model 2301 source, Best
Medical International, Springfield, VA). The energy distribution of
gammas exiting the capsule is mainly composed of three emission lines
(around 4.47, 4.9 and 27.4\,keV). In that sense, the brachytherapy
example is complementary to the linac examples.  Two files of 2.5\,GB
were generated, each containing $\num{1.04e8}$ gammas, described by 7
parameters each (energy, position, direction). One file was used to
train the GAN and the other one for evaluation.

For the examples considered in this work, the quantity of clinical
interest is the accuracy of the dose calculation. For the linac tests,
MC simulations were run in order to score the deposited energy in a
$20\times20\times20$ cm$^3$ water box with voxels of $4 \times 4
\times 4$ mm$^3$ (Elekta) and $2 \times 2 \times 2$ mm$^3$
(Cyberknife). As source, we used either a phase space file or
GAN-generated particles (PHPS$_{\textrm{GAN}}$).  $\num{e8}$ primary
photons were used for the Elekta and $\num{4e7}$ for the
Cyberknife. Particles in the phase space files were not used multiple
times. For the brachytherapy test, 79 seeds were evenly placed in the
prostate region of a CT image, emitting a total of $\num{e8}$ gammas,
in one case taken from a phase space file and in the other generated
by the GAN. The deposited energy was scored in $2\times 2\times 2$
mm$^3$ voxels.

In all cases, the MC relative statistical uncertainty $\sigma(k) =
{S(k)}/{\edep(k)}$ of the energy deposited in a voxel was computed
with the history-by-history method~\cite{Chetty2007}, with $S(k)$ the
statistical uncertainty in voxel with index $k$
(eq.~\ref{eq:stat_uncert}), $N$ the total number of primary events in
the simulation, $d_{k,i}$ the energy deposited in voxel $k$ by event
$i$, and $\edep(k)=\sum_i d_{k,i} $ the total deposited energy in
voxel $k$. The statistical uncertainty is the standard error of the
mean of the scored quantity, here the mean dose, and thus converges to
zero. The obtained values for the dose map in the three tests was less
than 3\% for all voxels with more than 10\% of the maximum dose.

\begin{eqnarray}
  \label{eq:uncertainty}
  S(k) & = & \sqrt{\frac{1}{N-1}\left( \frac{\sum_i^N d_{k,i}^2}{N} - \left(
             \frac{\sum_i^N d_{k,i}}{N}\right)^2 \right)}
             \label{eq:stat_uncert}
\end{eqnarray}


\begin{table}
  \centering
  \begin{tabular}[h]{ccc}
    PHSP & Size & Nb of particles \\\hline
    Elekta PRECISE 6MV & 2 files of 3.9 GB & $\num{1.3e8}$ photons each file\\
    CyberKnife IRIS 60mm & 2 files of 1.6 GB & $\num{5.8e7}$ photons each file  \\
    $^{125}$I brachy seed & 2 files of 2.5 GB & $\num{1.04e8}$ photons each file 
  \end{tabular}
  \caption{Characteristics of the three used datasets}
  \label{tab:datasets}  
\end{table}



In order to compare particles from a phase space file with those
generated by the GAN, marginal distributions of all parameters were
plotted. Simulations to compute the deposited energy in water using
phase space and GAN-generated particles were compared by analyzing the
voxel-by-voxel differences of the deposited energy. The distribution
of voxel differences naturally contains uncertainty. We evaluated the
similarity of this uncertainty between phase space and GAN generated
data. If the GAN produces realistic a phase space distribution, the
uncertainty should be similar in both cases. We thus compared the
distribution of voxel-wise differences between two simulations
performed with two different phase space files
($\Delta_{\textrm{PHPS}}$), and between two simulations using
particles from the GAN and a phase space file, respectively
($\Delta_{\textrm{GAN}}$). The differences were normalised by the
maximum value in the image, as a proxy for the prescribed dose,
denoted $\hat{\edep}_{\textrm{PHPS}_2} = \max_{\{k\}} D(k)$, see
equation~\ref{eq:diff}.




\begin{eqnarray}
  \Delta_{\textrm{PHPS}}(k) & = & \frac{\edep_{\textrm{PHPS}_2}(k) - \edep_{\textrm{PHPS}_1}(k)}{\hat{\edep}_{\textrm{PHPS}_2}}  \\
  \Delta_{\textrm{GAN}}(k) & = & \frac{\edep_{\textrm{PHPS}_2}(k) - \edep_{\textrm{GAN}}(k)}{\hat{\edep}_{\textrm{PHPS}_2}}
  \label{eq:diff}
\end{eqnarray}

Moreover, in every voxel, we computed the ratio between voxel
difference and uncertainty. If the statistical error were normally
distributed, the distributions of those ratios should have a zero mean
and unit standard deviation. Finally, for the linac experiments, we
computed depth dose curves (along $z$) and transversal dose profiles
at 20mm depth. For the brachytherapy example, we focused on the energy
distribution and visually inspected the distribution of deposited
dose.

\section*{Results}

\begin{figure}[h]
  \begin{center}
\includegraphics[width=0.99\linewidth]{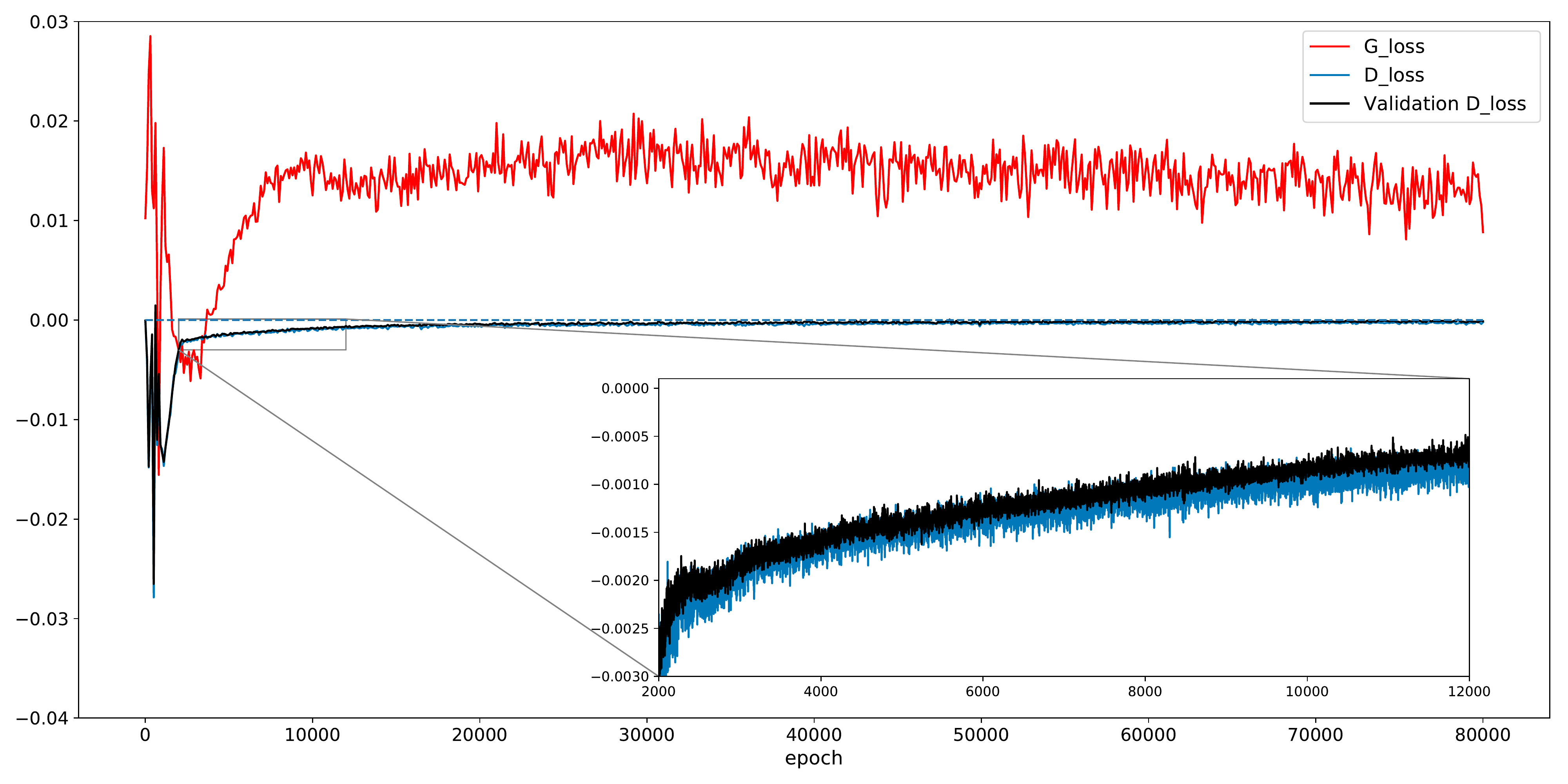}
    \caption{$J_D$ loss computed on training and validation datasets and $J_G$ as a function of iterations, for the Elekta test case (subsampled every 100 iterations, excepted in the magnified view).}
    \label{fig:loss}
  \end{center}
\end{figure}
\begin{figure}[h]
  \begin{center}
    \includegraphics[width=0.9\linewidth]{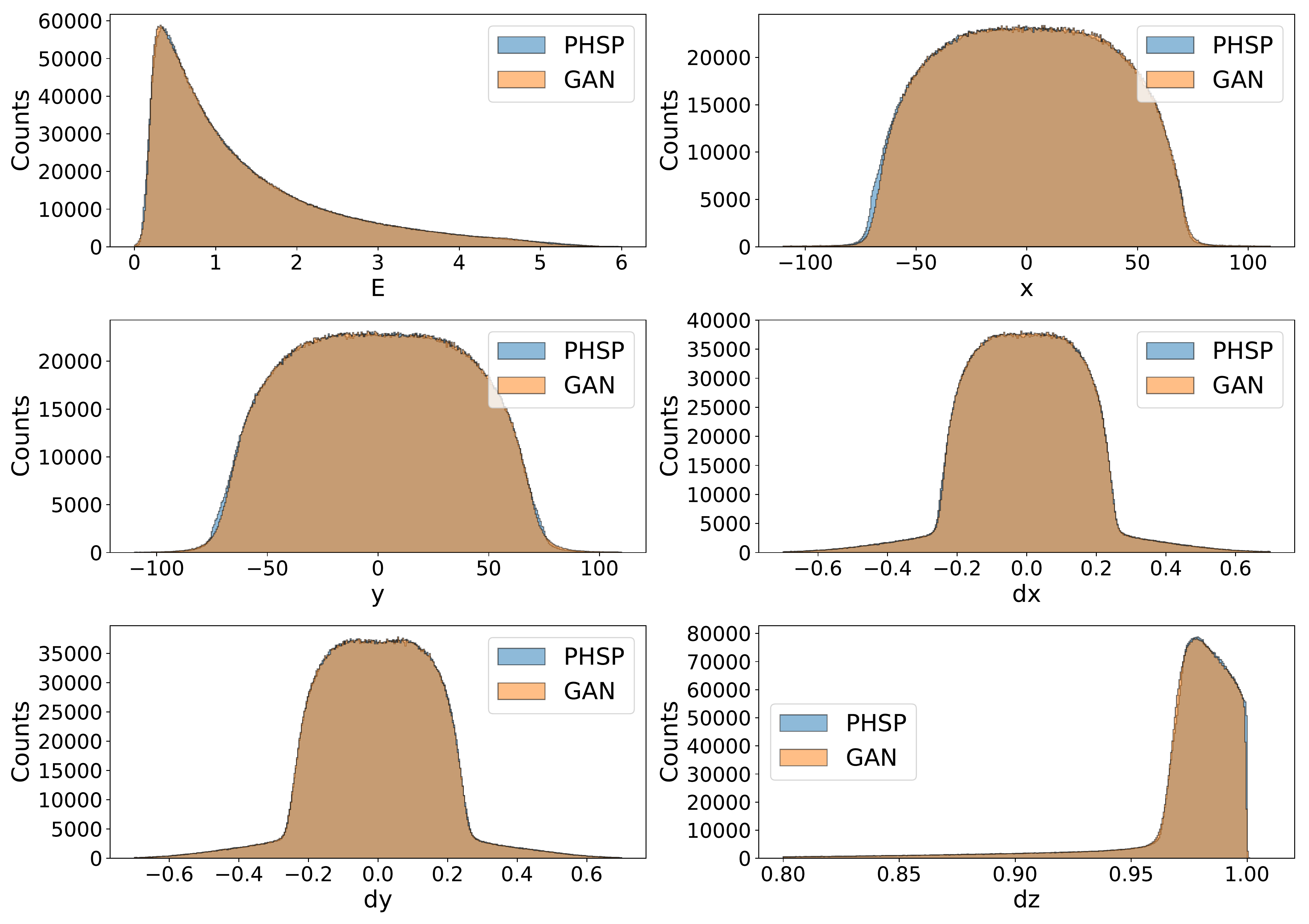}
    \caption{Marginal distributions of the 6 parameters obtained from
      the reference phase space file (PHSP) and from the GAN, for Elekta 6\,MV linac.}
    \label{fig:marginal_elekta}
  \end{center}
\end{figure}
\begin{figure}[h]
  \begin{center}
    \includegraphics[width=0.48\linewidth]{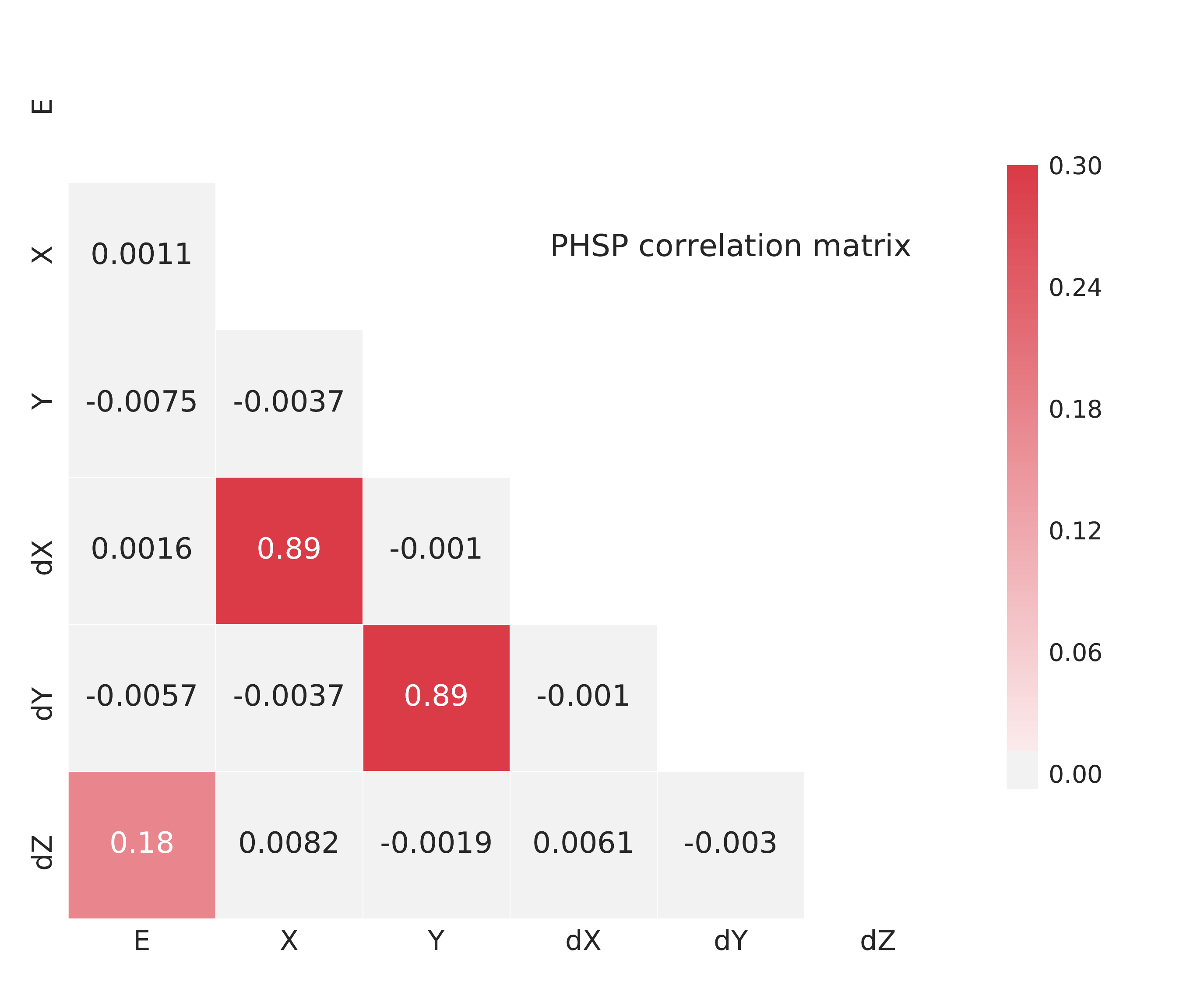}
    \includegraphics[width=0.48\linewidth]{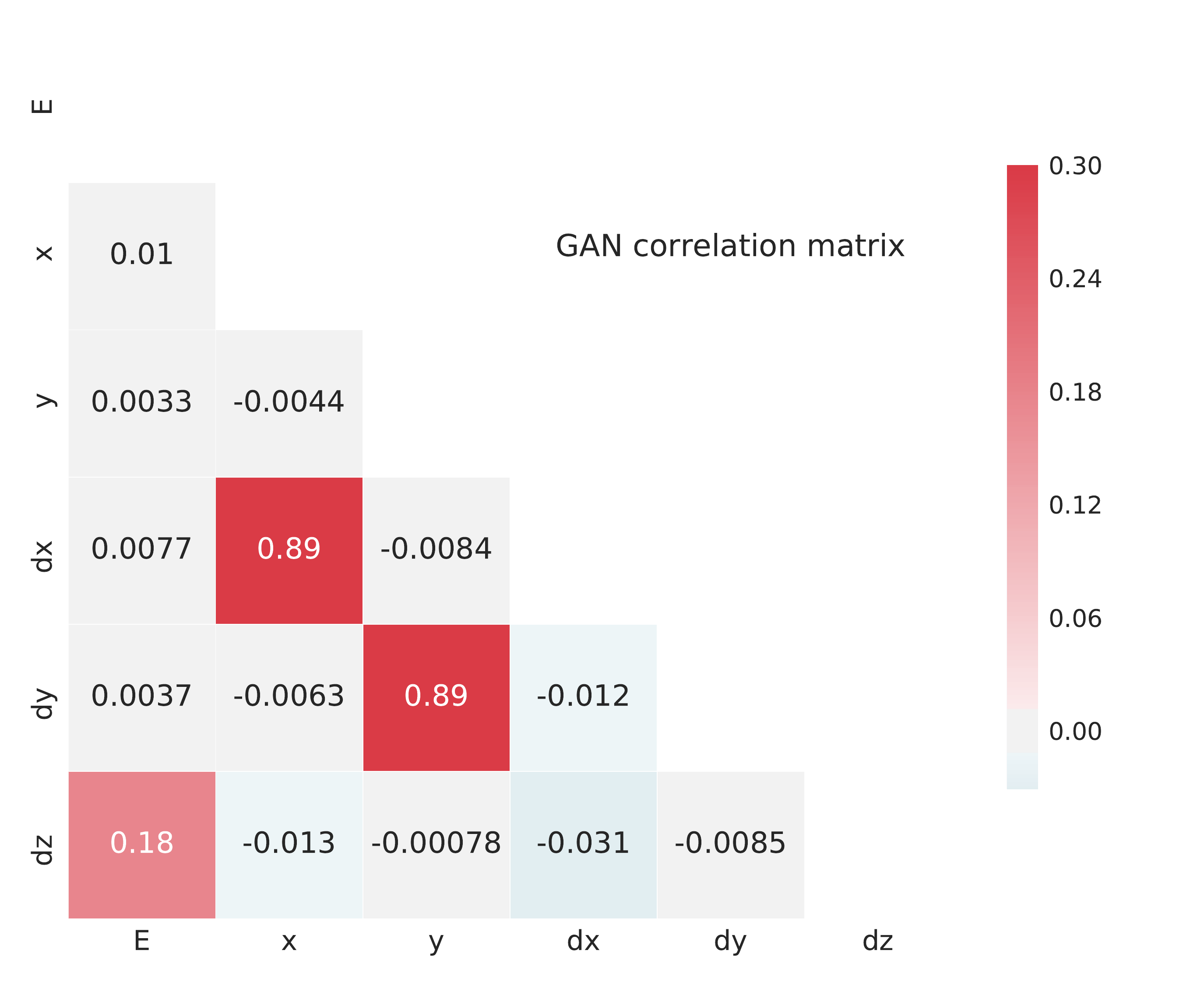}
    \caption{Correlation matrices between all 6 parameters for phase space file (PHSP, left) and GAN (right), for Elekta 6\,MV linac.}
    \label{fig:corr_matrix}
  \end{center}
\end{figure}
\begin{figure}[h]
  \begin{center}
    \includegraphics[width=0.9\linewidth]{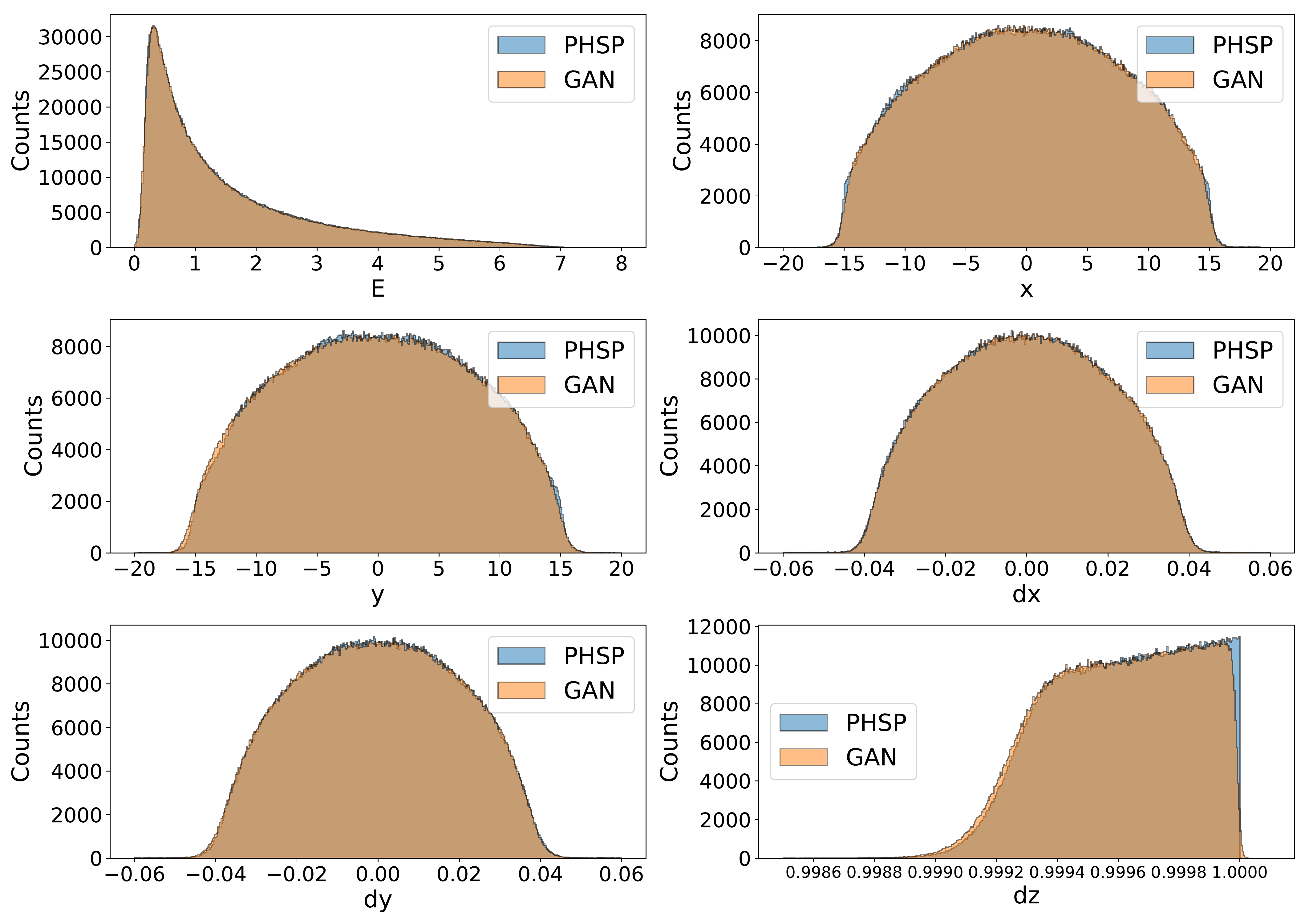}
    \caption{Marginal distributions of the 6 parameters obtained from
      the reference phase space file (PHSP) and from the GAN, for the Cyberknife linac with
      IRIS collimator.}
    \label{fig:marginal_ck}
  \end{center}
\end{figure}
\begin{figure}[h]
  \begin{center}
    \includegraphics[width=0.48\linewidth]{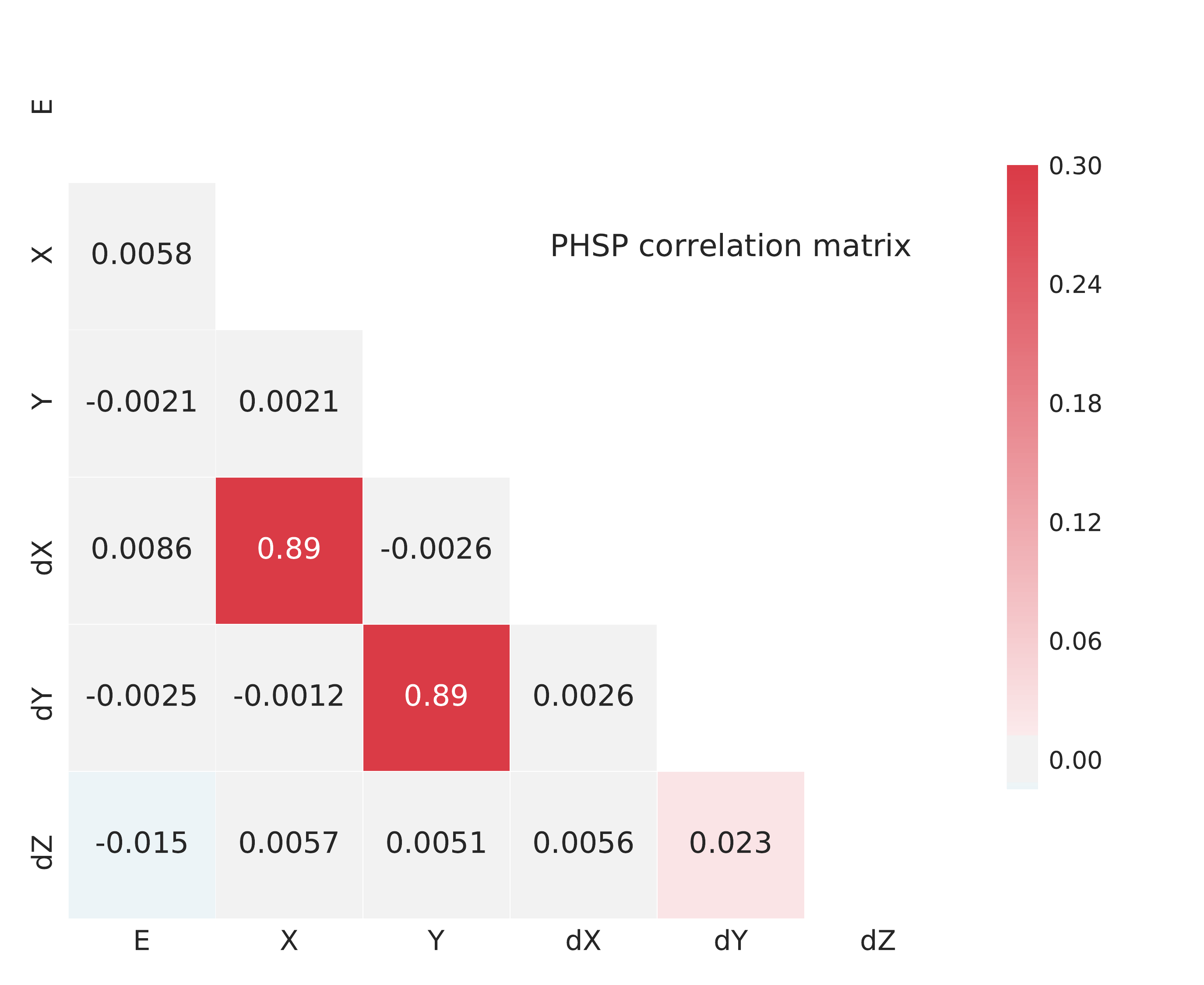}
    \includegraphics[width=0.48\linewidth]{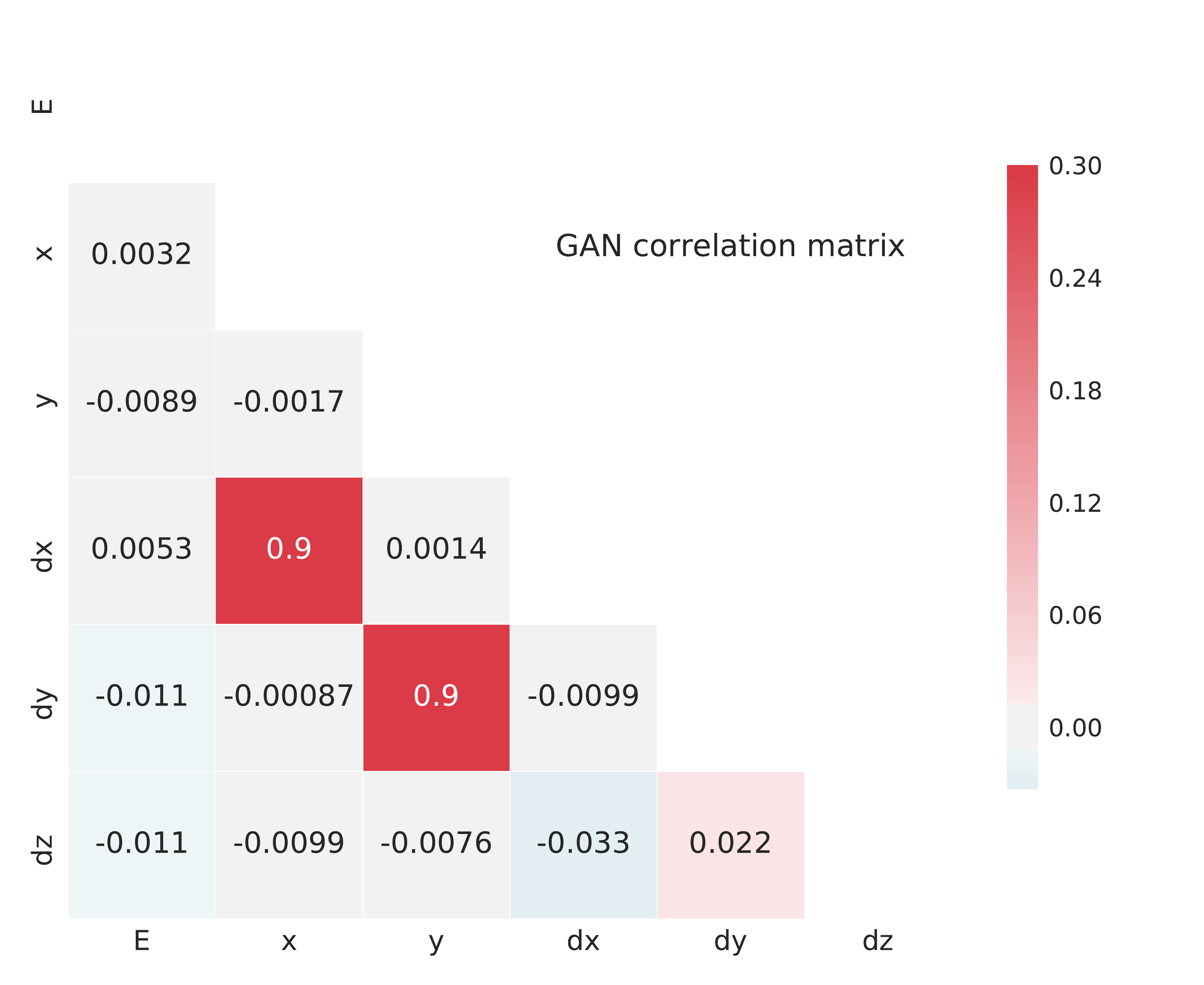}
    \caption{Correlation matrices between all 6 parameters for the phase space file 
      (PHSP, left) and GAN (right), for Cyberknife linac.}
    \label{fig:corr_matrix_ck}
  \end{center}
\end{figure}
\begin{figure}[h]
  \begin{center}
    \includegraphics[width=0.8\linewidth]{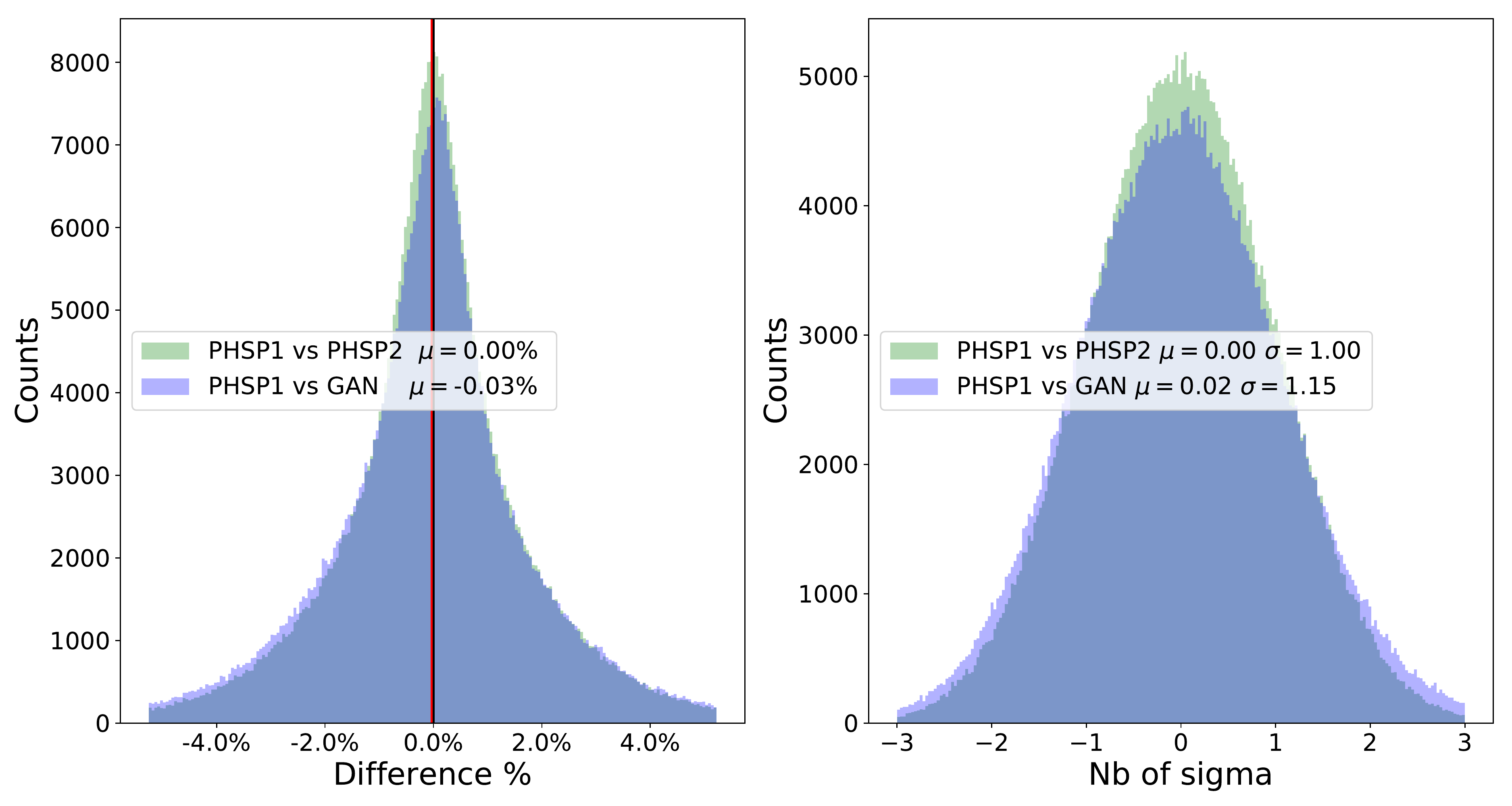}
    \includegraphics[width=0.8\linewidth]{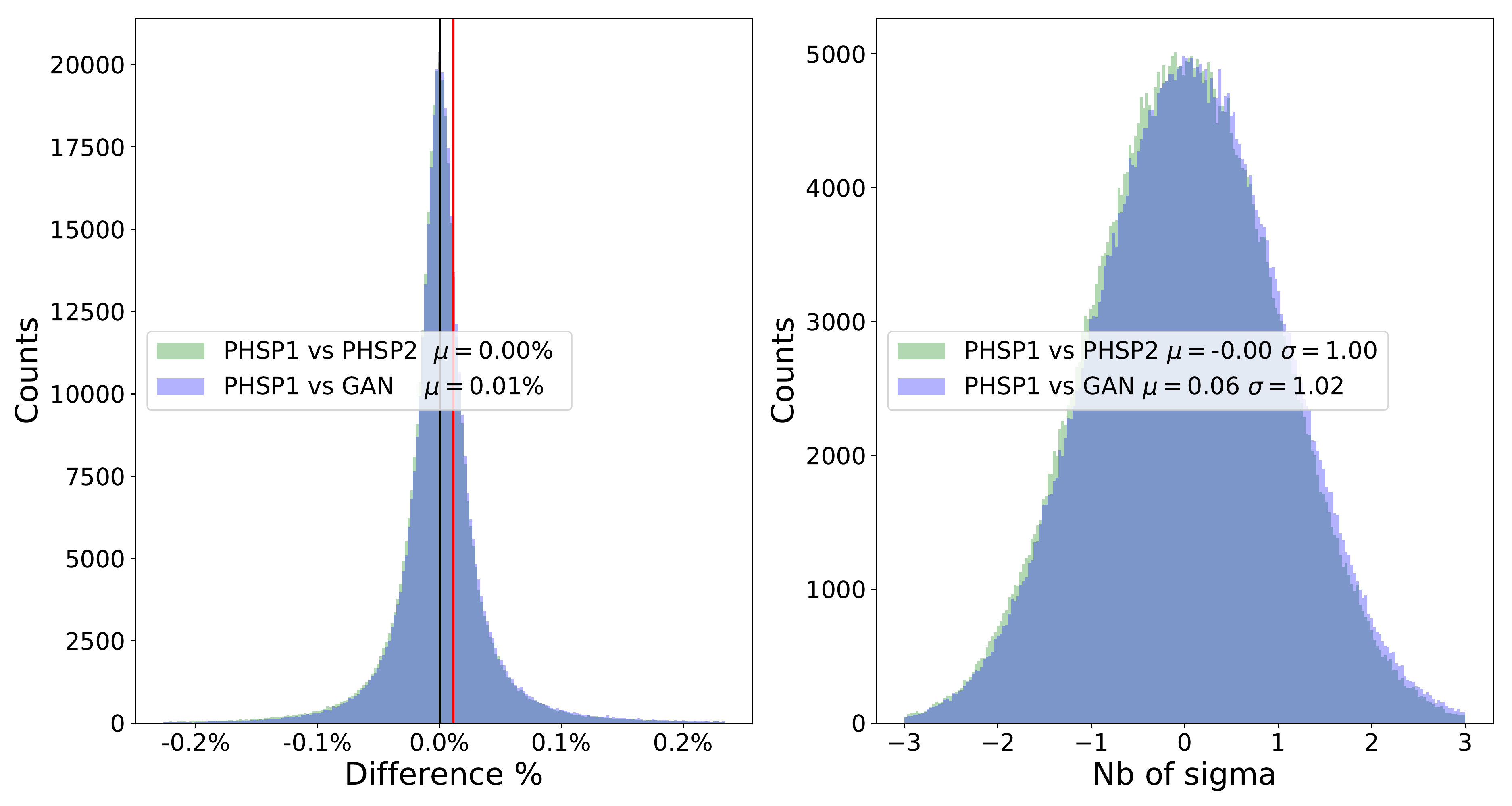}
    \includegraphics[width=0.8\linewidth]{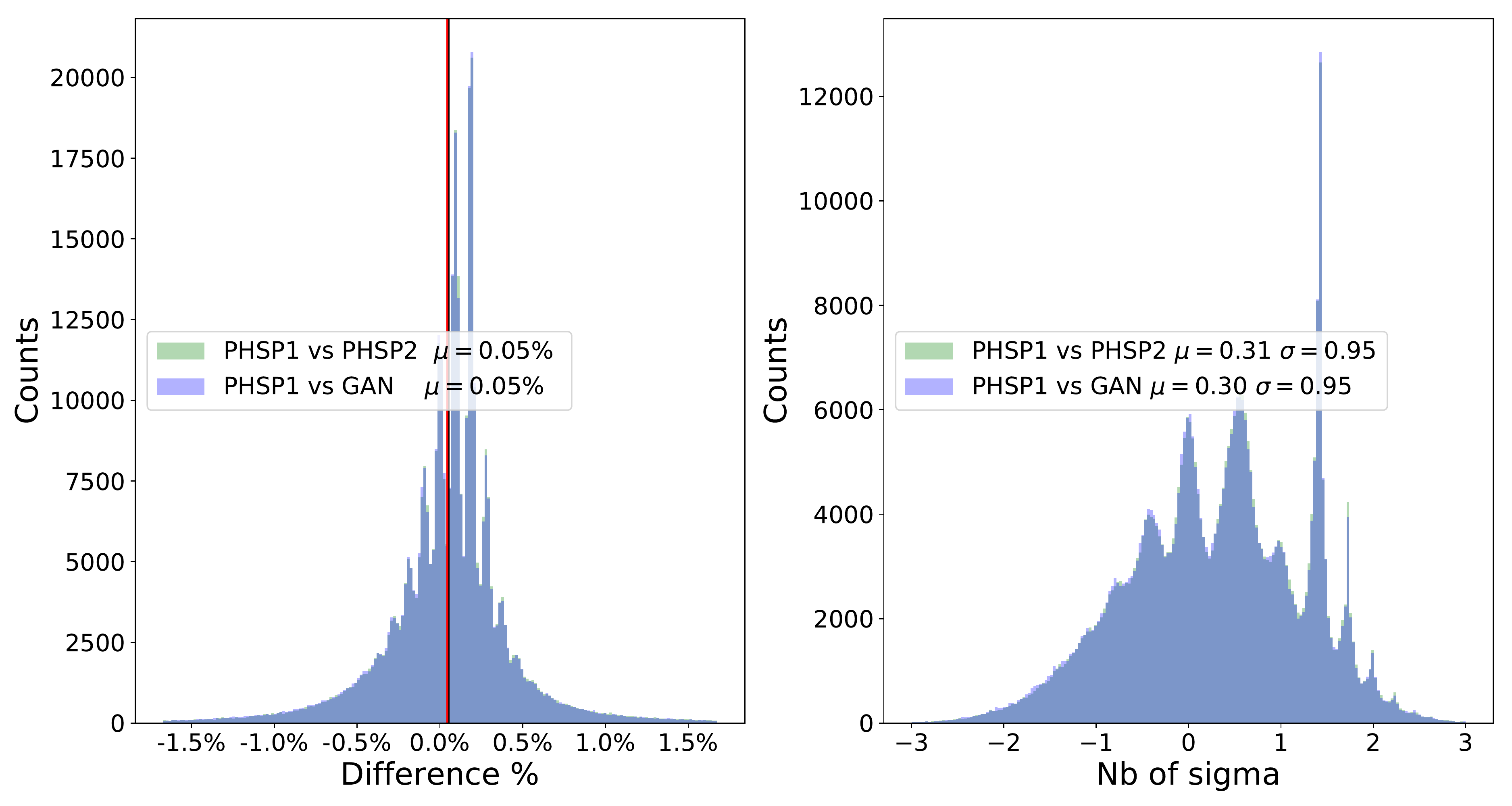}

    \caption{Distributions of relative differences between PHSP1 and
      PHSP2 and between PHSP1 and GAN. Vertical lines indicate the mean
      differences (in black for PHSP1-PHSP2 and in red for PHSP1-GAN). Right
      images show the difference relative to the statistical uncertainty,
      this distribution should have zero mean and standard deviation of one.
      Top row for Elekta machine, center row for Cyberknife device, and
      bottom row for brachytherapy test. For this last plot, the peaks
      structure correspond to the seeds in the CT image.}
    
    \label{fig:error_distrib}
  \end{center}
\end{figure}
\begin{figure}[h]
  \begin{center}
    \includegraphics[width=0.85\linewidth]{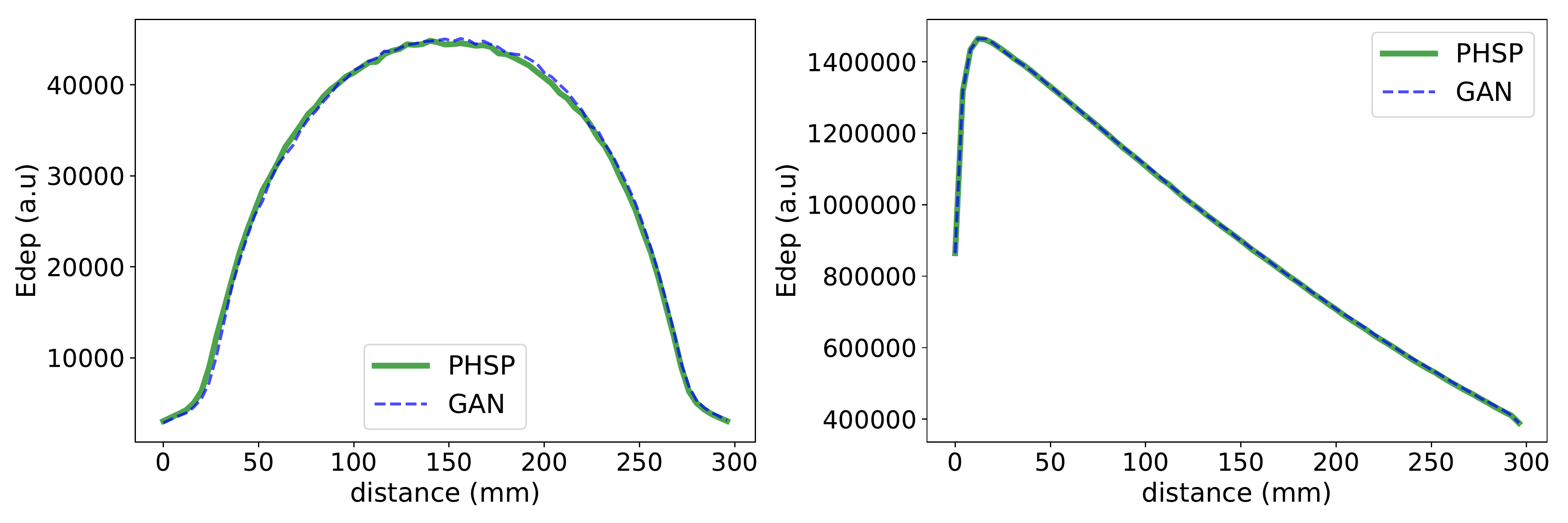}
    \includegraphics[width=0.85\linewidth]{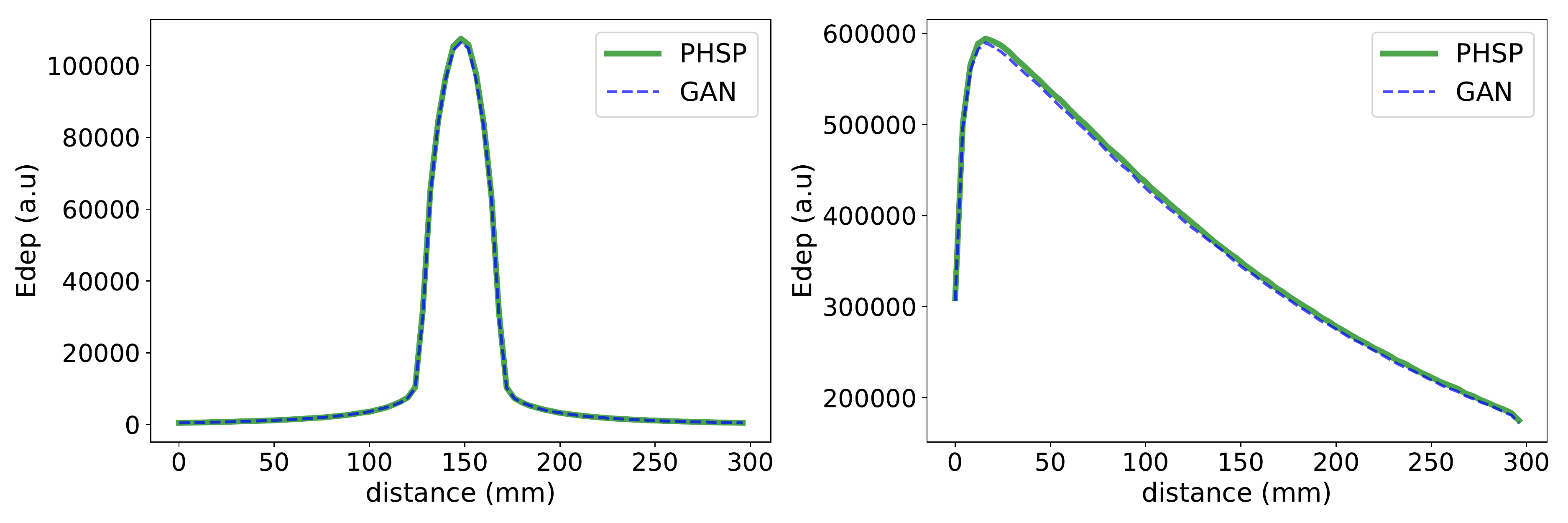}
    \caption{Transverse profiles at 20mm depth (left) and depth
      profiles (right) deposited energy for Elekta (top) and
      Cyberknife (bottom) machines. Curves were obtained for PHSP2 and
      GAN-based simulations.}
    \label{fig:profiles}
  \end{center}
\end{figure}
\begin{figure}[h]
  \begin{center}
\includegraphics[width=0.85\linewidth]{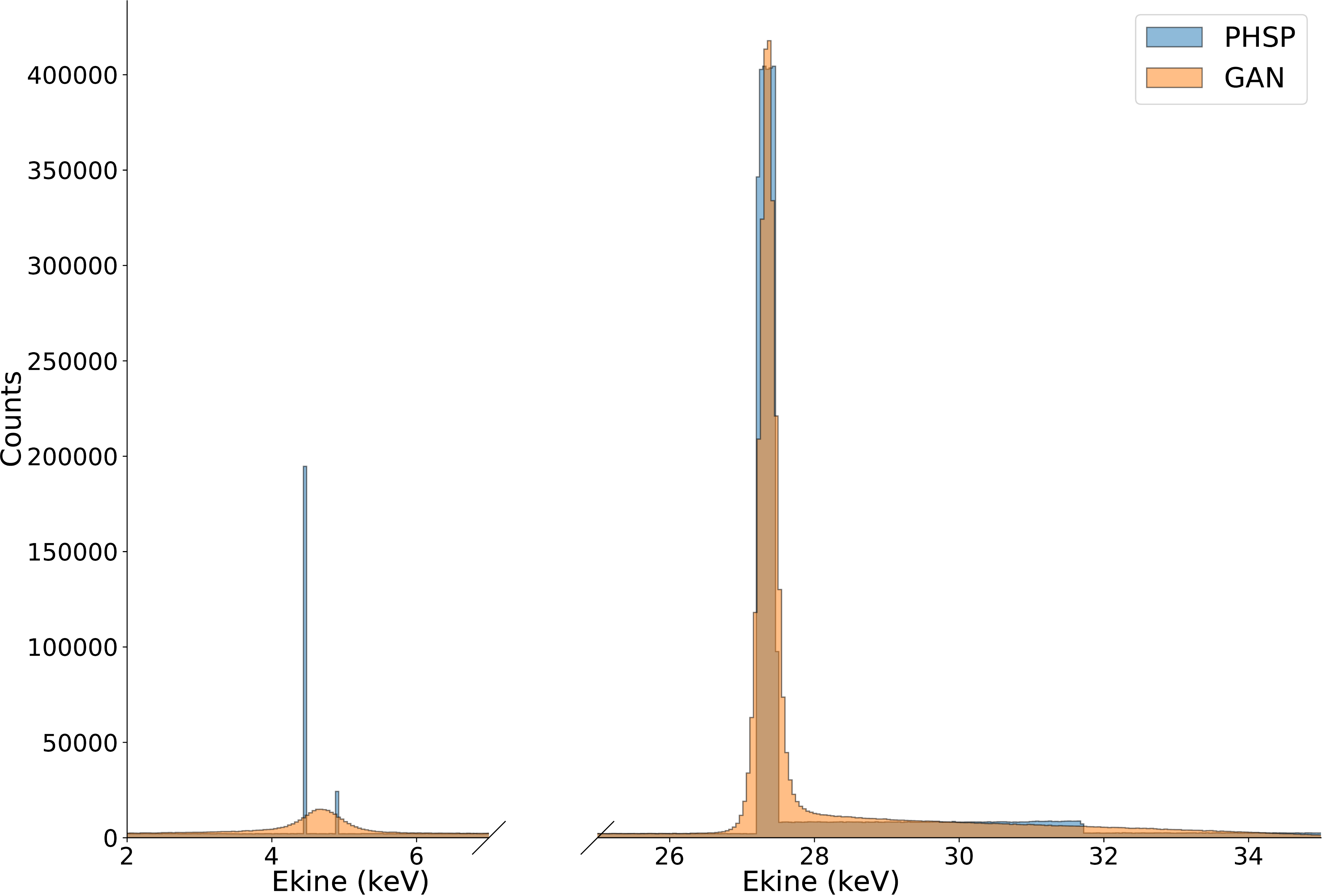}
    \caption{Closeup on the energy distribution obtained from the
reference brachytherapy phase space (PHSP) and from the GAN.}
    \label{fig:marginal_brachy}
  \end{center}
\end{figure}
\begin{figure}[h]
  \begin{center}
    \includegraphics[width=0.32\linewidth]{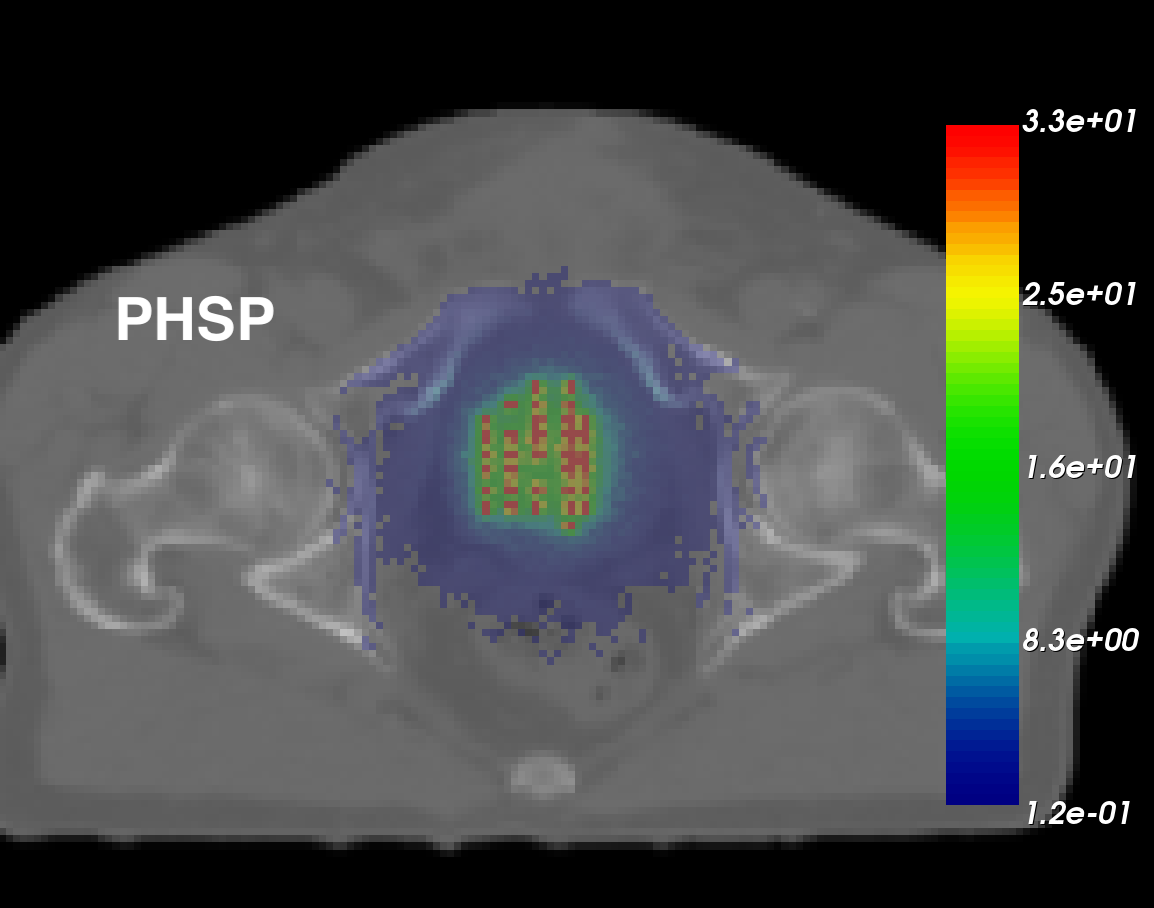}
    \includegraphics[width=0.32\linewidth]{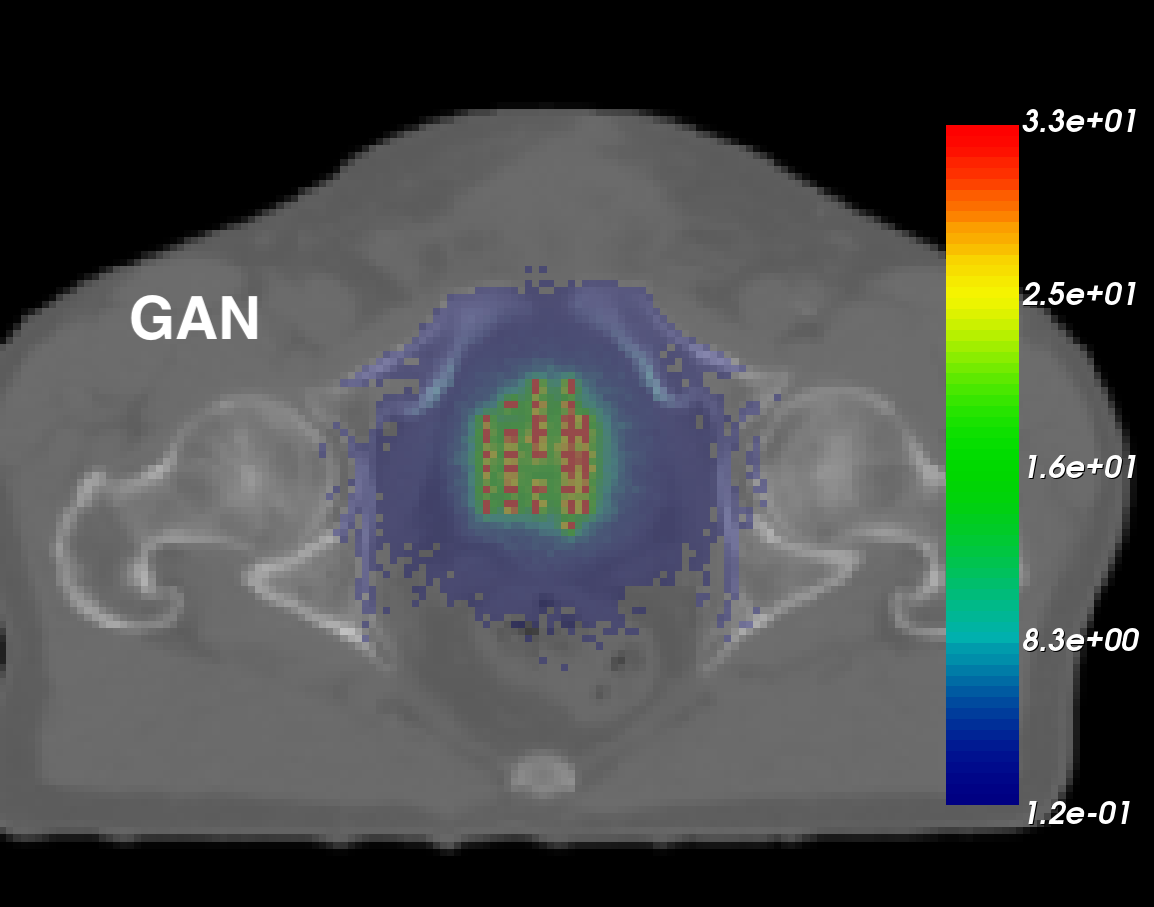}
    \includegraphics[width=0.32\linewidth]{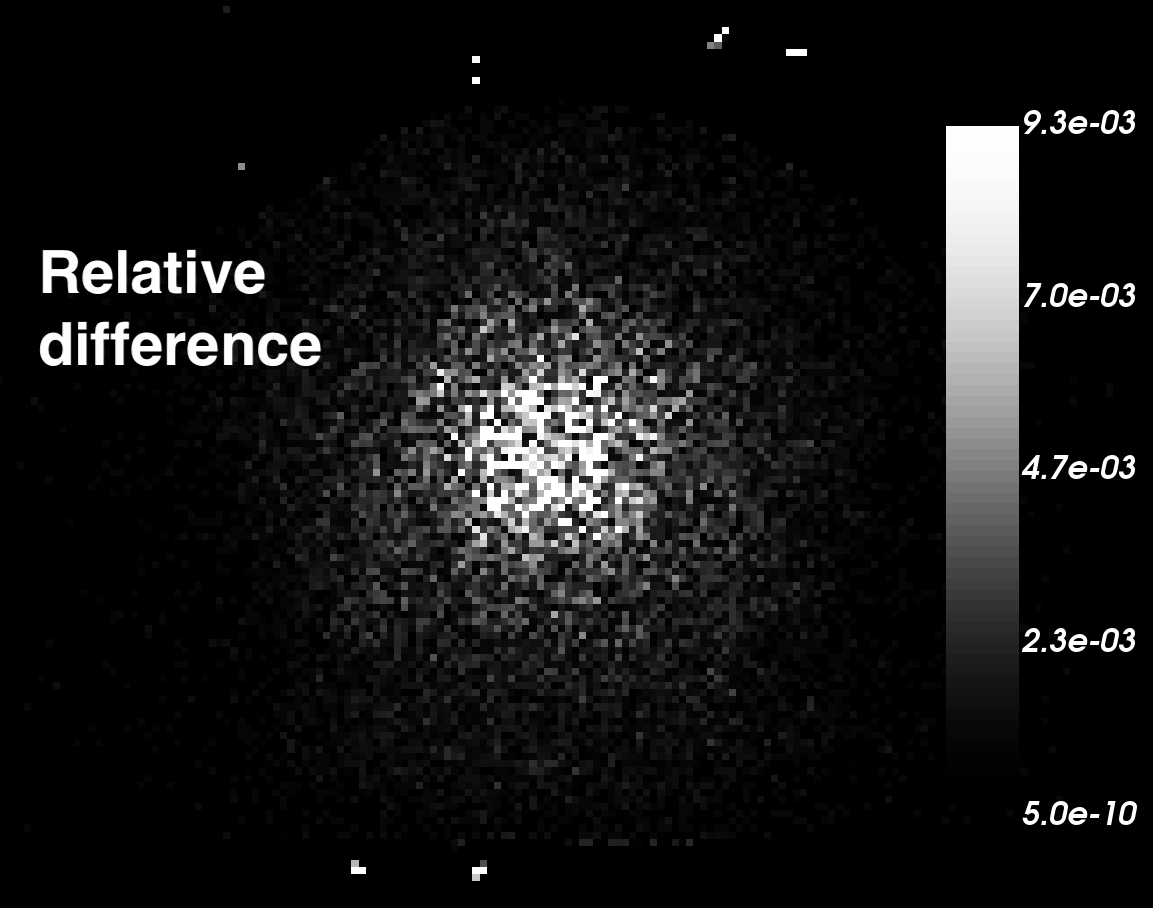}
    
    \caption{Slices of CT prostate image with deposited energy overlay
(in MeV), computed by phase space (PHSP, left) and GAN generated
(center) particles. The right image shows the dose difference
$\Delta_{\textrm{GAN}}$ relative to the maximum dose (the maximum
difference was below 4\%).}
    
    \label{fig:brachy_slices}
  \end{center}
\end{figure}

As an example, figure~\ref{fig:loss} depicts the evolution of the loss
$J_{D}$ on training and validation datasets and $J_G$ as a function of
iterations ($J_G$ does not depend on the kind of dataset). For visual
clarity, we only depict data for the Elekta test case, subsampled to
every 100 iterations (not subsampled in the magnified view).  $J_D$
(equation~\ref{eq:JDW}) is negative at the beginning because the
generator is not yet sufficiently trained, so that $D$ applied to real
data is larger than D applied to generated data. Once $G$ is trained,
$J_D$ is close to zero. Furthermore, $J_D$ is slightly less negative
when evaluated on the validation dataset rather than training dataset
because $D$ retains the validation dataset as slightly less
likely. The G-loss $J_G$ depicted larger variation than the other
losses but we did not investigate this further. Similar behavior was
obtained for the other tests. Figures~\ref{fig:marginal_elekta}
and~\ref{fig:marginal_ck} display the marginal distributions of the 6
parameters ($E,x,y,dx,dy,dz$; $z$ was fixed) extracted from the phase
space file compared with the ones obtained from the
GAN. Figure~\ref{fig:marginal_brachy} displays a closeup of the energy
distribution. Note that the datasets used for training the GAN were
always different from the ones used for
validation. Figures~\ref{fig:corr_matrix} and~\ref{fig:corr_matrix_ck}
also illustrate the correlations (covariance normalised by the product
of their standard deviations) between the 6 parameters for the two
linac tests. The left panel in figure~\ref{fig:error_distrib} shows
the distribution of the relative differences $\Delta_{\textrm{PHSP}}$
and $\Delta_{\textrm{GAN}}$ for all three tests. The mean differences
are indicated with vertical lines. The right hand panels show the
distribution of the ratio between differences and uncertainty, which
should ideally depict a mean value of zero and a standard deviation of
one. Figures~\ref{fig:profiles} show transversal and depth profiles of
deposited energy for both tests. Figures~\ref{fig:brachy_slices}
illustrate the deposited energy obtained for the brachytherapy test
case from simulations with particles from phase space files and
through GAN, respectively. The training process took around 2 hours
for all tests and generation of $\num{e6}$ samples from the GAN took
about one second. The final GAN model requires less than 10 MB of
storage space.


Training GAN is notoriously difficult: the models may not converge and
mode collapses are common~\cite{Arjovsky2017}. This was also our
experience and, in the following, we summarize our observations
regarding the hyperparameters of the training process.

\begin{itemize}
\item We observed that the Wasserstein version of the GAN is
  required. With the conventional GAN formulation, we did not achieve
  acceptable results.

\item The number of neurons ($H=400$) and layers (3 hidden layers for
  each G and D) were empirically set. Fewer neurons (around 300) lead
  to inferior results, while a larger number of neurons took more
  time to converge and did not seem to improve results.

\item The dimension of the latent variable $\bz$ is representative of the
underlying latent space structure of the multidimensional
distribution. In this work, it was therefore fixed to 6 for
linac tests and 7 for the brachytherapy test. We observed few differences
with 5 or 7 dimensions. However, a too low value (lower than 5) lead
to degraded results.



\item According to~\cite{Arjovsky2017}, it is suggested to perform
  more discriminator updates than generator updates. Here, we
  selected four discriminator updates for one generator update. A lower
  ratio (2:1) degraded results.

\item The learning rate was set to $\num{e-5}$ and we did not find any
clear improvement with other values. The batch size was empirically
set to $\num{e4}$. Smaller numbers of samples per batch lead to
inaccurate probability densities in the generated distributions (which
are 6/7-dimensional). This is probably because the multidimensional
phase space is too sparsely represented by such samples to be
faithfully learned. Larger batches did not really improve the
results. Figure~\ref{fig:loss} shows a typical training process. The
values of the loss function strongly oscillated during the first few thousands of
iterations and then tended to slowly converge and oscillate around a
fixed value. Repeated training with the same set of parameters lead to
slightly different results because of the stochastic nature of the
learning process.

\end{itemize}


\clearpage

\section*{Discussion}


The distributions generated by the GAN were close to those represented
the phase space files as shown in
figures~\ref{fig:marginal_elekta},~\ref{fig:corr_matrix},~\ref{fig:marginal_ck},
and~\ref{fig:corr_matrix_ck}. Note that the figures only display
marginal distributions and correlation between pairs of parameters,
but high order correlations were also modeled by the GAN. In
particular, X-dX and Y-dY correlations are related to the cone
geometry of the linac beam.

When the GAN was used to compute deposited energy in a water box,
overall results were good and could not be distinguished visually or
by looking at the transversal or depth profiles
(figure~\ref{fig:profiles}). Still, differences between phase space
and GAN based simulation results are slightly larger than differences
between two simulations using phase space files. For example, with the
Elekta linac, the distributions of differences show a small shift
towards negative values, visible in the top-left panel of
figure~\ref{fig:error_distrib}. This shift is however not
dosimetrically relevant (less than 0.04\% of the maximum dose). A
similar effect is also visible in the distribution of the differences
relative to the statistical uncertainty. The top and bottom right
panels in figure~\ref{fig:error_distrib} depict zero mean and unity
standard deviation for $\Delta_{\textrm{PHSP}}$, while small shifts
(0.02 and 0.06) are present for $\Delta_{\textrm{GAN}}$. We also
verified that the difference distribution between two simulations,
both using GAN generated particles, was very close to the one between
two phase space based simulations (not shown in the figure because
curves completely overlapped with the $\Delta_{\textrm{PHSP}}$
distribution). The last plots in figures~\ref{fig:error_distrib}
and~\ref{fig:brachy_slices} show the almost unnoticeable differences
between the deposited energy distributions. As a conclusion, the
experiments illustrate that a GAN generated particle distribution is
close to the distribution represented by the phase space file, but not
exactly equivalent: differences between GAN and phase space files are
larger than between two phase space files. The difference can hardly
be seen in the deposited energy here, but may be important for other
applications.

For the brachytherapy case, figure~\ref{fig:marginal_brachy} displays
a close-up around the main energy peaks, illustrating how the GAN only
approximates the peak values. It seems that sharp features in the
distributions are still difficult to model with this
method. Nevertheless, the dose maps were comparable. If an exact
representation of the energy spectrum is required, a workaround could
consist in training several GANs: one for each peak (ignoring the
energy parameter) and one for the continuous part. We note that for
the Elekta linac, the 511\,keV peak represents less than 0.4\% of the
total number of photons.

As already mentioned in papers related to GAN~\cite{Goodfellow2014,
Arjovsky2017}, training a GAN is still a difficult process,
involving manual tuning of hyperparameters and empirical decisions. In
this work, the same set of hyperparameters, given in
section~\ref{sec:parameters}, led to consistent results for all three
tested phase space files. The hyperparameter space is large and other more optimal
hyperparameters than the ones used in the presented work might
exist. Pruning techniques may also be employed in order to trim the
network size by removing unimportant neurons to improve speed
and results. Moreover, better results may be achieved with
a different network architecture or training process. Findings described
here may serve as a first guide to further investigations.

Since the initial GAN proposal~\cite{Goodfellow2014}, numerous
variants, among them WGAN, have been investigated in the literature
with more than 500 papers per month during the end of
2018~\cite{ganzoo2018}. Further works are still needed to evaluate the
interest of those variants, in particular to investigate if the
difficulty to precisely model sharp features of the distributions that
have been described earlier could be overcome. Also transfer learning
may be well adapted here: a first network trained for a given phase space
could be used as a starting point for the training of another one.
Note also that GANs are usually employed for problems with a higher
number of dimensions and less smooth distributions (natural images,
speech) than phase space data. Other methods, such as Gaussian mixture
models~\cite{Doerner2016}, may also be useful to model phase space files.

Overall, the proposed approach has several advantages. It allows
modelling a large file of several GB by one of about 10 MB (486406
neuron weights in the G model). The generation of particles from a GAN
is a very fast process (1 second for $\num{e6}$ particles) and an
arbitrary number of particles can be generated from the GAN while the
phase space dataset is finite. The use of a generator instead a phase
space file also greatly simplifies the simulation workflow.  The
infrastructure for training a GAN on a new phase space is generic and
convenient to setup. Users may only need to set hyperparameters
values.  The learning process in this work involved only $\num{e8}$
particles, which means that computation time is preserved compared to
a larger phase space. Further investigations might explore how
accurately a GAN can be trained based on a smaller training dataset.


\section{Conclusion}

In this work, we proposed a GAN architecture to learn large phase
space distributions used in Monte Carlo simulations of linacs and
brachytherapy. The advantages of this approach lie in the compact size
of the model, the ability to quickly generate a large number of
particles, and the very generic nature of the process that may
potentially be applied to a large range of sources of
particles. Indeed, this approach could be extended to other types of
simulations where complex distributions of particles are involved. We
think that the exploratory work presented here is a first of a kind
involving advanced machine learning methods for Monte Carlo
simulations and can be applied to a large class of applications in
medical physics.








\section*{Acknowledgments}

This work was performed within the framework of the SIRIC LYriCAN
Grant INCa-INSERM-DGOS-12563, and the LABEX PRIMES (ANR-11-LABX-0063)
of Universit{\'e} de Lyon, within the program “Investissements
d'Avenir” (ANR- 11-IDEX-0007) operated by the ANR. We gratefully
acknowledge the support of NVIDIA Corporation with the donation of the
Titan Xp GPU used for this research.

\section*{References}
\bibliographystyle{my_bibstyle.bst}
\bibliography{all}

\end{document}